  \documentclass[prl,aps,twocolumn,showpacs]{revtex4}
\usepackage[dvips]{graphicx}
\usepackage{amsmath,  amssymb, array, graphics,  amsfonts}
\usepackage{dcolumn}
\begin{document}
\topmargin=-15mm

\title {\bf
 A SOLUTION TO THE PROBLEM OF THE SKIN EFFECT WITH A DISPLACEMENT
CURRENT IN
THE MAXWELL PLASMA BY THE METHOD OF EXPANSION IN EIGENFUNCTIONS}

\author{\bf Yu.F. Alabina, A. V. Latyshev, and A. A. Yushkanov}

\affiliation  {Department of Mathematical Analysis and Department of
Theoretical Physics, Moscow State Regional University,  105005,
Moscow, Radio st., 10--A}

\begin{abstract}
A problem of the skin effect in the Maxwell plasma is solved
analytically by the method of expansion in eigenfunctions based on
the Vlasov–-Maxwell kinetic equation with a self-consistent electric
field. Specular electron reflection from the boundary is used as a
boundary condition.

   \end{abstract}


\date{\today}
\maketitle

\section{I. Introduction}
The skin effect is caused by the electron gas response to an
external variable electromagnetic field tangential to the surface
[1]. This classical problem has been studied by many authors (for
example, see [1–3]). The present work develops an analytical method
of solving boundary problems for systems of equations describing the
behavior of electrons and an electric field in the half-space of
weakly ionized plasma. This method is extremely convenient, because
it allows the sought-after distribution function to be derived in an
explicit form. The method being developed is based on the idea of
expansion of the solution in generalized singular eigenfunctions of
the corresponding characteristic system [2] obtained after variable
separation. A solution to the characteristic system in the space of
generalized functions [4] gives eigenfunctions with a continuous
spectrum covering the entire positive real semiaxis. The structure
of the discrete spectrum is elucidated by finding zeros of the
dispersion function, and eigenfunctions of this spectrum are
determined. A general solution to the system of the Vlasov-–Maxwell
equations is constructed based on solutions for continuous and
discrete spectra. The proof of the expansion in the eigenfunctions
is reduced to a solution of the integral equation with the Cauchy
kernels. The last is reduced to the Riemann boundary problem in the
theory of functions of complex variables. The solvability conditions
and the Sokhotskii formulas allow all unknown expansion coefficients
in the solution of the initial boundary problem to be calculated.
Let us assume that the Maxwell plasma occupies the half-space $x>0$
, where $x$ is the coordinate orthogonal to the plasma boundary. Let
the external electric field has only one y-component. Then the
self-consistent electric field inside the plasma will also have only
one $y$-component $E(x)e^{-i\omega t}$. We now consider the kinetic
equation for the electron distribution function:
$$
\dfrac{\partial f}{\partial t}+\text{v} _{x}\dfrac{\partial
f}{\partial x}+eE(x)e^{-i\omega t}\dfrac{\partial f}{\partial
p_y}=\nu(f_0-f(t,x,\mathbf{v})). \eqno{(1)}
$$
where $\nu$ is the frequency of electron collisions with ions, $e$
is the electron charge, and $f_0(\nu)$ is the Maxwell equilibrium
distribution function:
$$
f_0(\text{v})=n\left(\dfrac{\beta}{\pi}\right)^{3/2}
\exp(-\beta^2\text{v}^2),\quad \beta=\dfrac{m}{2k_BT}.
$$
Here $k$ is the Boltzmann constant, $T$ is the plasma temperature,
$\nu$ is the electron velocity, $m$ is the electron mass, and $n$ is
the electron concentration.

Let us assume that the field strength is such that the linear
approximation is applicable. Then the distribution function can be
represented in the form
$$
f=f_0\left(1+C_y\exp(-i\omega t)h(x,\mu)\right),
$$
where $\textbf{C}=\sqrt{\beta}\text{v}$ is the dimensionless
velocity of electron and $\mu=C_x$.  We now introduce dimensionless
quantities $t_1=\nu t, \quad x_1=\nu\sqrt{\beta}x$, and
$$
e(x_1)=\dfrac{\sqrt{2}e}{\nu\sqrt{mk_BT}}E(x_1).
$$
Then we will
write again $x$ instead of $x_1$ . In the new variables, kinetic
equation (1) and the field equation with allowance for the
displacement
current are written as follows:
$$
\mu\dfrac{\partial h}{\partial x}+z_0\,h(x,\mu)=e(x),
 \eqno{(2)}
$$
$$
e''(x)+Q^2e(x)=-i\dfrac{\alpha}{\sqrt{\pi}}
\int\limits_{-\infty}^{\infty}\exp(-{\mu'}^2)\,h(x,\mu')\,d\mu',\eqno{(3)}$$$$
\quad Q=\dfrac{\omega l}{c},\quad z_0=1-i\omega\tau,
$$
where $l$ is the free path of the electron,
$\delta=\dfrac{c^2}{2\pi\omega\sigma_0}$, $\delta$ is the classical
depth of the skin layer, $\sigma_0=\dfrac{e^2n}{m\nu}$, $\sigma_0$
is the electric conductance, $\alpha=\dfrac{2l^2}{\delta^2}$,
$\alpha$ is the anomaly parameter.

Let us formulate conditions for the distribution function and field
on the plasma boundary:
$$
h(0,\mu)=h(0,-\mu), \quad 0<\mu<+\infty, \quad e(0)=e_s. \eqno{(4)}
$$

We search for a distribution function and field that decay with
increasing distance from the surface:
$$
h(+\infty,\mu)=0, \qquad -\infty<\mu<+\infty,  \qquad  e(\infty)=0.
\eqno{(5)}
$$

Without loss of generality, we further set $e_s=1$.

\section{II. Eigenfunctions and eigenvalues}

Separation of variables (see [2])
$$
h_\eta(x,\mu)=\exp(-z_0\dfrac{x}{\eta})\Phi(\eta,\mu), $$$$
e_\eta(x)=\exp(-z_0\dfrac{x}{\eta})E(\eta),
$$
where $\eta$ is a complex spectral parameter, reduces system of
equations (2) and (3) to the characteristic system
$$
(\eta-\mu)\Phi(\eta,\mu)=\dfrac{\eta}{z_0}E(\eta), \eqno{(6)}
$$
$$
\left[z_0^2+Q
 ^2\eta^2\right]E(\eta)=-\dfrac{i\alpha \eta^2}{\sqrt{\pi}}n(\eta),
$$
where
$$
n(\eta)=\int\limits_{-\infty}^{\infty}e^{-\mu^2}
\Phi(\eta,\mu)d\mu.\eqno{(7)}
$$

 From Eqs. (6) and (7) we find the
eigenfunctions of the continuous spectrum in the class of
generalized functions [3]:
$$
\Phi(\eta,\mu)=\dfrac{a}{\sqrt{\pi}}\eta^3e^{-\eta^2}
P\dfrac{1}{\eta-\mu}+ \lambda(\mu)\delta(\eta-\mu), \eqno{(8)}
$$
$$
E(\eta)=\dfrac{az_0}{\sqrt{\pi}}\eta^2e^{-\eta^2}, \quad
a=-i\dfrac{\alpha}{z_0^3}. \eqno{(9)}
$$

Taking into account the decrease of the distribution function and
electric field far from the boundary, the positive real semiaxis $0
<x<+\infty$ is taken to mean the continuous spectrum of the boundary
problem. The eigenfunctions of the continuous spectrum
$h_\eta(x,\mu)$ and $e_\eta(x)$ are decreasing functions of the
variable $x$ for $\Re z_0>0$. The eigenfunctions in equalities (8)
and (9) have been normalized by the condition
$$
\int\limits_{-\infty}^{\infty}e^{-\mu^2}\Phi(\eta,\mu)d\mu=
\left[1+\left(\dfrac{\omega l}{c}\right)^2\eta^2\right]e^{-\eta^2},
$$
and the dispersion function
$$
\lambda(z)=1+\left(\dfrac{Q}{z_0}\right)^2z^2+\dfrac{az^3}{\sqrt{\pi}}
\int\limits_{-\infty}^{\infty} \dfrac{e^{-\mu^2}d\mu}{\mu-z},
$$
has been introduced.

Let us designate $b=\dfrac{Q^2}{z_0^2}$ and express the dispersion
function of the problem in terms of the dispersion function of the
Van Kampen plasma $\lambda_0(z)$:
$$
\lambda(z)=1+(b-a)z^2+az^2\lambda_0(z),
$$
where
$$
\lambda_0(z)=\dfrac{1}{\sqrt{\pi}}\int\limits_{-\infty}^{\infty}
\dfrac{\mu e^{-\mu^2}d\mu}{\mu-z}.
$$

For the dispersion function in the vicinity of the point at
infinity, the asymptotic expansion
$$
\lambda(z)=(b-a)^2+(1-\dfrac{a}{2})-\dfrac{3a}{4z^2}-\dfrac{15a}{8z^4}-...,
\qquad z\rightarrow\infty,
$$
is fulfilled.

We now elucidate the structure of the discrete spectrum by the
method developed in [2, 3]. By definition, this spectrum consists of
zeros of the dispersion function laying outside of the cut
$(-\infty, \infty)$.

Let $N$ be the number of zeros. Since the dispersion function has a
double pole at the point $z =\infty$, the number of its zeros is
$$
N=2+\dfrac{1}{2\pi}[\arg \lambda(z)]_{\gamma_\varepsilon},
\eqno{(10)}
$$
 where $\gamma_\varepsilon$ is a contour passing
clockwise over the cut $(-\infty,\infty)$ at distance $\varepsilon$
and having no zeros inside.

Taking the limit in Eq. (10) when $\varepsilon\rightarrow 0$ , we
obtain
$$
N=2+\dfrac{1}{2\pi}\left[\arg
\dfrac{\lambda^+(\tau)}{\lambda^-(\tau)}\right]_{(-\infty,\infty)}=2+\dfrac{1}{\pi}\left[\arg
\dfrac{\lambda^+(\tau)}{\lambda^-(\tau)}\right]_{(0,\infty)}
$$

Here $\lambda^\pm(\tau)=\lambda(\mu)\pm i \pi a \mu^3 e^{-\mu^2}$
are the maximum and minimum values of the function $\lambda(z)$ in
the cut.

Let us consider the region $D^+$ (we designate by $D^-$ its external
boundary) in the $a$ plane whose boundary is set by the parametric
equations
$$
\partial D^+=\{\alpha=\alpha_1+i\alpha_2: \quad \Re
\lambda^+(\mu)=0, $$$$ \Im \lambda^+(\mu)=0, \quad
-\infty<\mu<\infty\}.
$$

By analogy with [2], we can demonstrate that 1) if $a\in D^+,  N =
4$ and 2) if $a \in D^-, N = 2$. The mode with $a \in \partial D$ is
not considered here, since it has already been studied in detail in
[3].

Let us write down (discrete) eigenfunctions corresponding to the
obtained discrete spectrum $\{\pm \eta_k: \lambda(\eta_k)=0, k=0,
1\}$:
$$
\Phi(\eta_k, \mu)=\dfrac{a}{\sqrt{\pi}}\sum_{k=0}^1\dfrac{\eta_k^3
e^{-\eta_k^2}}{\eta_k-\mu}, $$$$
E(\eta_k)=\dfrac{az_0}{\sqrt{\pi}}\sum_{k=0}^1 \eta_k^2
e^{-\eta_k^2} \quad (k=0,1).
$$

We note that in the last formulas, $k = 0$ when $a \in D^-$ and
$k=0,1$ when $a\in D^+$.

\section{III. Analytical problem solution}

Let us represent the general solution of system (2)-–(5) in the form
of expansion in eigenfunctions of the discrete and continuous
spectra, automatically satisfying the boundary conditions at
infinity:
$$
h(x,\mu)=\dfrac{a}{\sqrt{\pi}}\sum\limits_{k=0}^{1}
\dfrac{A_k\eta_k^3}{\eta_k-\mu}\exp\Big(-\eta_k^2-
\dfrac{z_0x}{\eta_k}\Big)+
$$
$$
+ \int\limits_{0}^{\infty}\exp\Big(-\dfrac{z_0x}{\eta}\Big)
A(\eta)\Phi(\eta,\mu)\,d\eta, \eqno{(11)}
$$
$$
e(x)=\dfrac{az_0}{\sqrt{\pi}}\sum\limits_{k=0}^{1}
A_k\eta_k^2\exp\Big(-\eta_k^2-\dfrac{z_0x}{\eta_k}\Big)+
$$
$$
+\dfrac{az_0}{\sqrt{\pi}}\int\limits_{0}^{\infty}
\exp\Big(-\eta^2-\dfrac{z_0x}{\eta}\Big)\eta^2\,A(\eta)\,d\eta.
\eqno{(12)}
$$

Here $ A_k \quad (k= 0, 1)$ are unknown coefficients of the discrete
spectrum with $A_1=0$ for $a \in D^-, A(\eta) $ is unknown function
called the coefficient of the continuous spectrum, $\Re(z_0/\eta_k )
>0 \quad  (k = 0, 1)$, and $\Re z_0 = 1$.

Substituting expansions (11) and (12) into the boundary conditions,
we obtain the following integral equations:
$$
a\varphi(\mu)+\int\limits_0^\infty A(\eta)\Phi(\eta,\mu)d\eta-
\int\limits_0^\infty A(\eta)\Phi(\eta,-\mu)d\eta=0,\eqno{(13)}
$$
$$
\dfrac{1}{\sqrt{\pi}}\sum\limits_{k=0}^{1}A_k\eta_k^3\exp(-\eta_k^2)+$$$$+\dfrac{1}
{\sqrt{\pi}}\sum\limits_{k=0}^{1}\eta^2\exp(-\eta^2)A(\eta)d\eta=\dfrac{1}{az_0}.\eqno{(14)}
$$
where
$$
\varphi(\mu)=\dfrac{1}{\sqrt{\pi}}
\sum\limits_{k=0}^{1}A_k\eta_k^3\exp(-\eta_k^2)
\Big(\dfrac{1}{\eta_k-\mu}-\dfrac{1}{\eta_k+\mu}\Big).
$$

Let us transform Eq. (14) setting $A(-\eta)=-A(\eta)$, that is,
expanding the coefficient $A(\eta)$ to the entire real axis as an
odd one. Considering that $\Phi(-\eta,-\mu)=\Phi(\eta,\mu)$, we
reduce Eq. (13) to the form
$$
\varphi(\mu)+\int\limits_{-\infty}^\infty
A(\eta)\Phi(\eta,\mu)d\eta=0, \quad -\infty<\mu<\infty,
$$
or after substitution of the eigenfunctions into this equation,
$$
\dfrac{a}{\sqrt{\pi}}\int\limits_{-\infty}^{\infty}\dfrac{\eta^3A(\eta)
\exp(-\eta_k^2)\,d\eta}
{\eta-\mu}+\lambda(\mu)A(\mu)+$$$$+a\varphi(\mu)=0, \quad
-\infty<\mu<\infty. \eqno{(15)}
$$

Let us introduce the auxiliary function
$$
N(z)=\dfrac{1}{\sqrt{\pi}}\int\limits_{-\infty}^{\infty}
\dfrac{\eta^3\exp(-\eta^2)A(\eta)}{\eta-z}d\eta,
$$
whose boundary values, according to the Sokhotskii formulas, obey
the equality
$$
N^+(\mu)-N^-(\mu)=2\sqrt{\pi}i \mu^3
\exp(-\mu^3)A(\mu)=$$$$=\dfrac{A(\mu)}{a}[\lambda^+(\mu)-\lambda^-(\mu)].
$$

With the help of boundary values of the auxiliary function $N(z)$
and the dispersion function, we reduce the integral equation with
Cauchy's kernel (15) to the Riemann boundary problem
$$
\lambda^+(\mu)[N^+(\mu)+\varphi(\mu)]=\lambda^-(\mu)[N^-(\mu)+\varphi(\mu)],
$$
whose general solution has the form
$$
N(z)=-\dfrac{1}{\sqrt{\pi}}\sum_{k=0}^1A_k\eta_k^3\exp(-\eta_k^2)\times
$$
$$
\times\left[
\dfrac{1}{\eta_k-z}-\dfrac{1}{\eta_k+z}\right]+\dfrac{C_1z}{\lambda(z)}.
\eqno{(16)}
$$

Eliminating the first-order poles at points $\eta_k$, we obtain
$$
C_1=-\dfrac{1}{\sqrt{\pi}}A_k\eta_k^2\exp(-\eta_k^2)\lambda'(\eta_k)
\quad (k=0,1).
$$

Substituting general solution (16) into the Sokhotskii formula, we
obtain the coefficient for the continuous spectrum:
$$
\eta^2
\exp(-\eta^2)A(\eta)=\dfrac{C_1}{2\sqrt{\pi}i}\left[\dfrac{1}{\lambda^+(\eta)}
-\dfrac{1}{\lambda^-(\eta)}\right].
$$

We now return to Eq. (14) and write it in the form
$$
-\dfrac{1}{\lambda'(\eta_0)}-\dfrac{1}{\lambda'(\eta_1)}+\dfrac{1}{2\pi
i}
\int\limits_0^\infty\left[\frac{1}{\lambda^+(\eta)}-\frac{1}{\lambda^-(\eta)}\right]d\eta=$$$$=
\frac{1}{az_0C_1}. \eqno{(17)}
$$

After integration of Eq. (17) by the methods of contour integration,
we transform the last equation and calculate first the constant
$C_1$:
$$
C_1=\dfrac{1}{az_0J(a)}, \quad
J(a)=\dfrac{1}{2\pi}\int\limits_{-\infty}^\infty\frac{d\tau}{\lambda(i\tau)}=
\dfrac{1}{\pi}\int\limits_0^\infty\frac{d\tau}{\lambda(i\tau)},
$$
and then constants $A_k$ with the help of Eq. (14): $A_k=
-\dfrac{\sqrt{\pi}\exp(\eta_k^2)}{az_0J(a)\eta^2_k\lambda'(\eta_k)}$
$(k=0,1)$.

To calculate the impedance, we consider the electric field
derivative
$$
e'(0)=az^2_0C_1\cdot$$$$\cdot\left[\dfrac{1}{\eta_0\lambda'(\eta_0)}+\dfrac{1}{\eta_1\lambda'(\eta_1)}
-\dfrac{1}{2\pi
i}\int\limits_0^\infty\left[\dfrac{1}{\lambda^+(\eta)}-\dfrac{1}{\lambda^-
(\eta)}\right]\dfrac{d\eta}{\eta}\right].
$$

To integrate this expression, we take advantage of the
representation
$$
\dfrac{1}{\lambda(z)}=\dfrac{1}{2\pi
i}\int\limits_{-\infty}^{\infty}\left[\dfrac{1}{\lambda^+(\eta)}-\dfrac{1}{\lambda^-
(\eta)}\right]\frac{d\eta}{\eta-z}-$$$$-\sum_{k=1}^{1}\dfrac{2\eta_k}{(\eta^2_k-z^2)\lambda'(\eta_k)}.
\eqno{(18)}
$$

From equality (18) for $z =0$ , we obtain
$$
I=-\sum_{k=0}^1\dfrac{2}{\eta_k\lambda'(\eta_k)}+\frac{1}{2\pi
i}\int\limits_{-\infty}^\infty\left[\dfrac{1}{\lambda^+(\eta)}-\dfrac{1}{\lambda^-
(\eta)}\right]\frac{d\eta}{\eta}.
$$

Taking into account the evenness of the integrand, we obtain
$$
\frac{1}{2\pi
i}\int_{-\infty}^{\infty}\left[\dfrac{1}{\lambda^+(\eta)}-\dfrac{1}{\lambda^-
(\eta)}\right]\frac{d\eta}{\eta}=\frac{1}{2}+\frac{1}{\eta_0\lambda'(\eta_0)}+
\frac{1}{\eta_0\lambda'(\eta_0)}.
$$

Now it is clear that the derivative of the electric field is
$e'(0)=\dfrac{az^2_0}{2}C_1=\dfrac{z_0}{2J(a)}$
 and the expression for the surface impedance is
$$
Z=\dfrac{8\pi i \omega l}{c^2\,z_0} \left[\dfrac{1}{\pi}
\int\limits_{0}^{\infty}\dfrac{d\tau}{\lambda(i\tau)}\right]^{-1}.
\eqno{(19)}
$$

Let us express all constants in Eq. (19) in terms of
$\gamma=\dfrac{\omega}{\omega_p}$ and
$\varepsilon=\dfrac{\nu}{\omega_p}$, where $\omega_p=\dfrac{4\pi n
e_0^2}{m}$ is the plasma frequency,
$b=\dfrac{\gamma^2}{(\varepsilon-i\gamma)^2}v_c^2$,
$a=-i\dfrac{\gamma}{(\varepsilon-i\gamma)^3}v^2_c$, and
$v_c=\frac{1}{v_c\sqrt{\beta}}$.

\begin{figure}[h]
\begin{center}
\includegraphics[width=8.5cm,height=4.5cm]{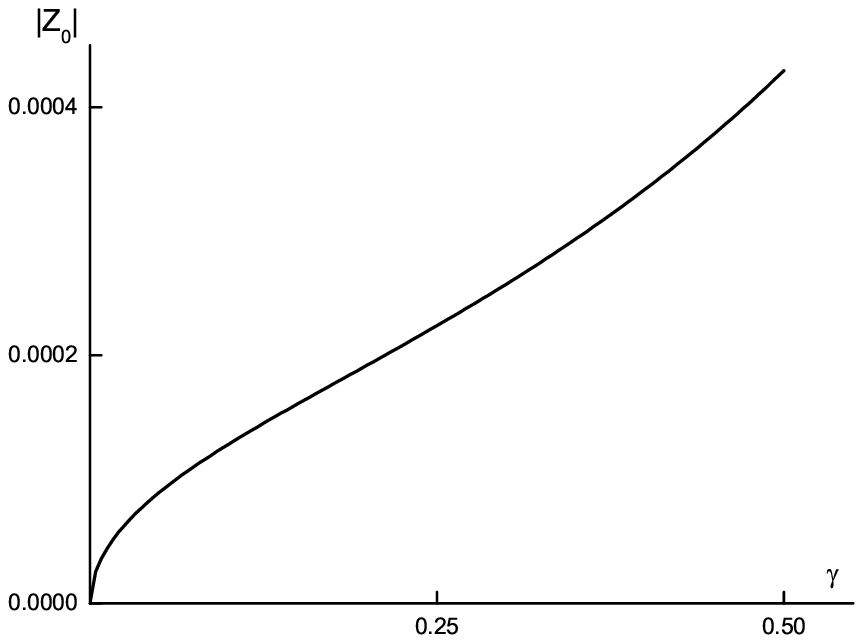}
\end{center}
\begin{center}
{Fig. 1. Modulus of the impedance.}
\end{center}
\end{figure}

\begin{figure}[h]
\begin{center}
\includegraphics[width=8.5cm,height=4.5cm]{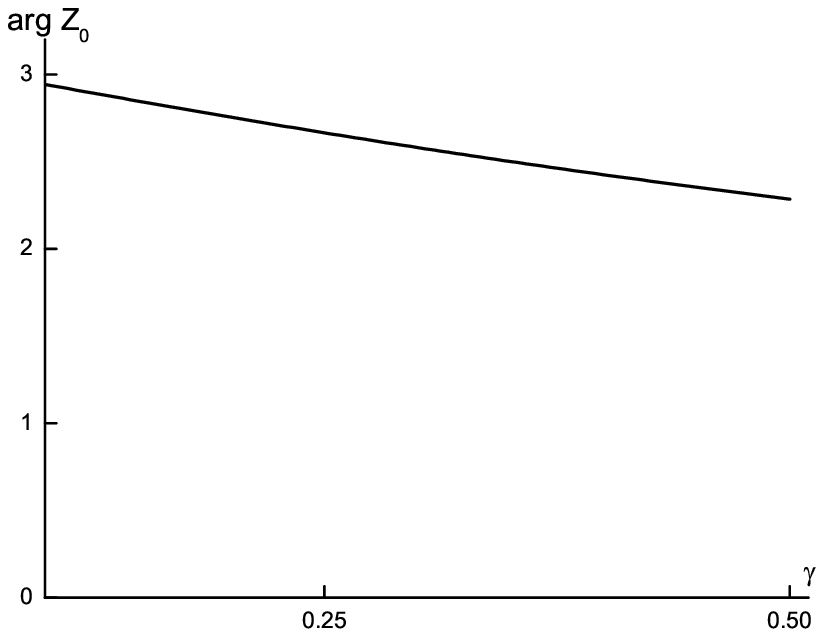}
\end{center}
\begin{center}
{Fig. 2. Argument of the impedance.}
\end{center}
\end{figure}

\begin{figure}[h]
\begin{center}
\includegraphics[width=8.5cm,height=4.5cm]{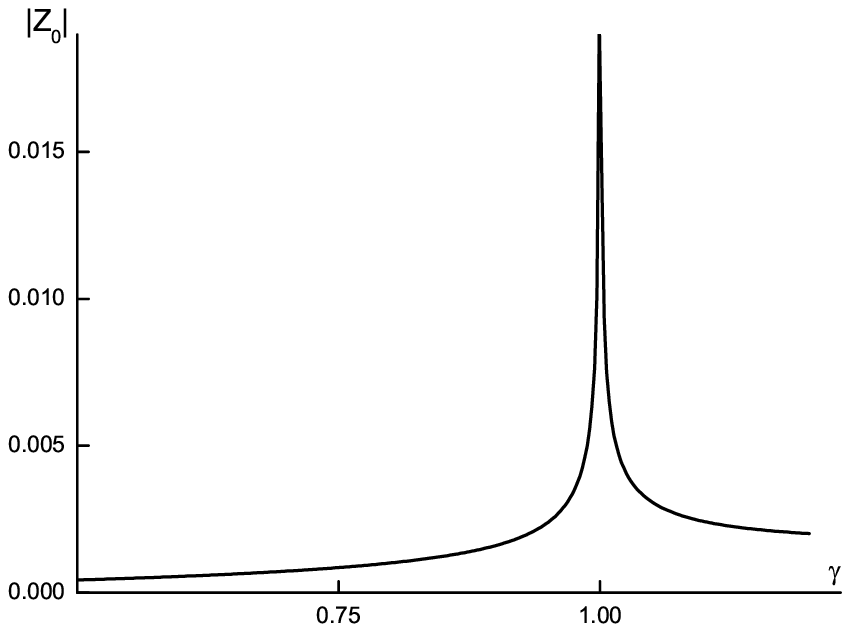}
\end{center}
\begin{center}
{Fig. 3. Modulus of the impedance.}
\end{center}
\end{figure}

\begin{figure}[t]
\begin{center}
\includegraphics[width=8.5cm,height=4.5cm]{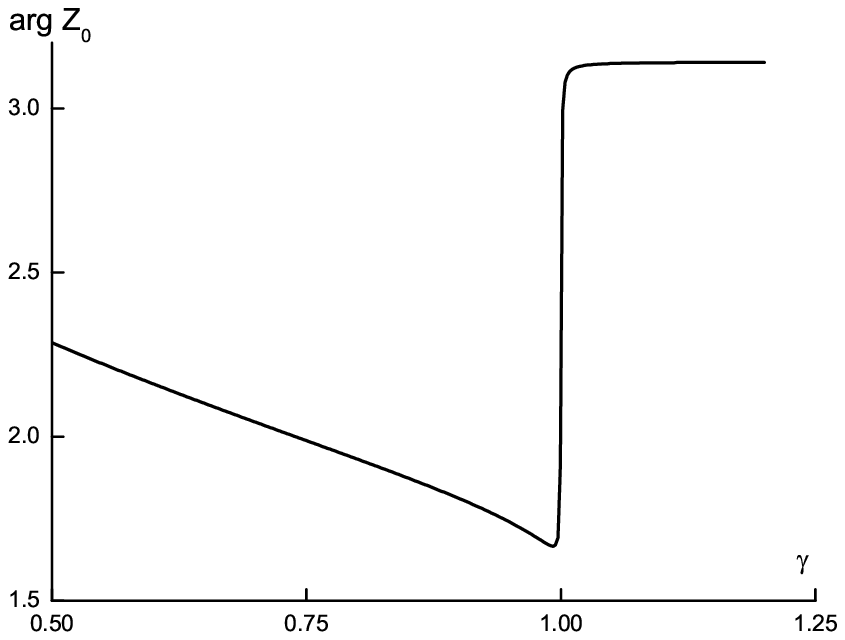}
\end{center}
\begin{center}
{Fig. 4. Argument of the impedance.}
\end{center}
\end{figure}

Let us now represent dispersion function (10) in the form
$$
\lambda(z)=1+\dfrac{\gamma^2v_c^2}{(\varepsilon-i\gamma)^2}z^2+i\dfrac{\gamma
v_c^2}{(\varepsilon-i
\gamma)^2}p(z)=$$$$=\dfrac{1}{(\varepsilon-i\gamma)^3}$$$$\left[(\varepsilon-i\gamma)^3+
(\varepsilon-i\gamma)\gamma^2v_c^2z^2+i\gamma v_c^2z^2+i\gamma
v_c^2p(z)\right],
$$
where
$p(z)=-\dfrac{z^3}{\sqrt{\pi}}
\int\limits_{-\infty}^\infty\dfrac{\exp(-\mu^2)}{\mu-z}d\mu.
$ After substitution of the expression obtained into formula (19)
for the surface impedance, we have
$$
Z=\dfrac{8\pi i \omega l}{c^2\,z_0}\times $$
$$\times\left[\dfrac{1}{\pi}
\int\limits_{0}^{\infty}\dfrac{(\varepsilon-i\gamma)^3dt}{(\varepsilon-i\gamma)^3+
(\varepsilon-i\gamma)\gamma^2v_c^2t^2+i\gamma
v_c^2p(t)}\right]^{-1}.
$$

\section{IV. Conclusions}

The expression for the impedance can be represented as $Z=RZ_0$,
where $R=2\pi \omega\delta c^{-2}$ is the magnitude of the normal
skin effect and $Z_0$ is the dimensionless impedance. The behavior
of the dimensionless impedance modulus is shown in Figs. 1 and 3,
and the behavior of its argument is illustrated by Figs. 2 and 4 for
$\varepsilon=10^{-3}$ and $v_c=10^{-3}$. The plots in Figs. 3 and 4
are drawn near the resonance, that is, when the parameter $\gamma$
passes through the value $\gamma=1$ for $\omega=\omega_p$.

 An
analysis of plots drawn in Figs. 1 -- 4 demonstrates that near the
plasma resonance, the modulus of the impedance has a sharp maximum
which is not observed in the low-frequency limit or in the theory of
normal skin effect, and the argument of the impedance changes
abruptly near the resonance.

\end{document}